\documentclass[aps, pra, twocolumn, showpacs,superscriptaddress]{revtex4-1}
\usepackage{graphicx,subfigure,braket}
\usepackage{amsmath,amssymb}
\usepackage{color}
\usepackage{natbib,hyperref}
\usepackage{placeins}
\usepackage[T1]{fontenc}
\begin{document}

\title{Robust shaped pulses for arrays of superconducting or semiconductor spin qubits with fixed Ising coupling}

\author{David W. Kanaar}
\affiliation{Department of Physics, University of Maryland Baltimore County, Baltimore, MD 21250, USA}
\author{J.~P.~Kestner}
\affiliation{Department of Physics, University of Maryland Baltimore County, Baltimore, MD 21250, USA}

\begin{abstract}
 A major current challenge in solid-state quantum computing is to scale qubit arrays to a larger number of qubits. This is hampered by the complexity of the control wiring for the large number of independently tunable interqubit couplings within these arrays. One approach to simplifying the problem is to use a qubit array with fixed Ising ($ZZ$) interactions. When simultaneously driving a specific subset of qubits in such a system, the dynamics are confined to a set of commuting $\mathfrak{su}$(2) subalgebras. Within these $\mathfrak{su}$(2)s we describe how to perform $X$-gates and $\frac{\pi}{2}$ $ZZ$ rotations robustly against either leakage, which is the main source of error in transmon qubits, or coupling fluctuations, which is the main source of infidelity in flux or semiconductor spin qubits. These gates together with virtual-$z$ gates form a universal set of gates for quantum computing. We construct this set of robust gates for two-edge, three-edge, and four-edge vertices, which compose all existing superconducting qubit and semiconductor spin qubit arrays. 
\end{abstract}

\maketitle

\section{Introduction}
Full-scale quantum computing requires a large number of logical qubits which consist of even more physical qubits combined through error-correcting codes. In order for these error-correcting codes to succeed at arresting errors, the infidelity of the physical qubit operations must be upper bounded by a threshold value, estimated to be between $10^{-4}$ and $10^{-2}$, depending on the specific code \cite{nielsen_quantum_2010,wang_surface_2011,fowler_surface_2012}. One- and two-qubit operations have reached this range of infidelities in many types of qubits, such as trapped ion \cite{harty_high-fidelity_2014,gaebler_high-fidelity_2016,ballance_high-fidelity_2016,erhard_characterizing_2019}, semiconductor electron spin \cite{yang_silicon_2019,yoneda_quantum-dot_2018,petit_universal_2020,noiri_fast_2022,mills_two-qubit_2022,huang_fidelity_2019} and superconducting qubits \cite{sheldon_procedure_2016,hong_demonstration_2020,weiss_fast_2022,moskalenko_high_2022}. However, it remains a challenge to scale up these systems to a large number of qubits while maintaining sub-threshold infidelities.

Scaling qubit arrays made from electron spins confined by quantum dots in semiconductor devices is promising because of the existing semiconductor industry \cite{maurand_cmos_2016,gonzalez-zalba_scaling_2021}. However, maintaining high fidelity in large semiconductor spin arrays is difficult in part because of charge noise, capacitive crosstalk between all of the barrier gates that control the exchange coupling, and variability in the dynamic range of the exchange between different pairs \cite{cifuentes_impact_2023}. Examples of existing arrays consist of 4 to 6 qubits in a linear chain or square array \cite{unseld_2d_2023,philips_universal_2022,lawrie_simultaneous_2023}. Meanwhile, the arrays of superconducting qubits from IBM \cite{jurcevic_demonstration_2021,kim_evidence_2023} and Google \cite{acharya_suppressing_2023} already contain on the order of a hundred qubits. However, this is still small compared to the estimated number of physical qubits needed for practical problems -- on the order of a million \cite{fowler_surface_2012}. 
Qubit arrays with tunable two-qubit coupling require calibration of each coupling \cite{arute_quantum_2019}. Additionally, they can suffer from residual exchange although this can be mitigated using modified driving techniques \cite{heinz_analysis_2023}.

In both the semiconductor spin qubit case and the superconducting qubit case, using fixed coupling arrays can help circumvent these difficulties while lowering fabrication overhead for coupling controls \cite{le_scalable_2023}. The trade-off to using fixed coupling is that implementing a universal set of gates with a coupling that cannot be turned off is not trivial. However, by driving a specific subset of qubits in a fixed Ising ($ZZ$) coupling array the Hamiltonian decomposes into a set of commuting $\mathfrak{su}$(2) subalgebras. Ref.~\cite{kanaar_non-adiabatic_2022} shows how to create a universal set of gates in a chain and honeycomb array of qubits in the absence of error using this decomposition. A universal set of gates robust against coupling and amplitude fluctuations for a honeycomb array of flux qubits has been numerically found \cite{le_scalable_2023} using a similar decomposition. This is a logical choice since the main source of infidelity in flux qubits is flux noise that causes fluctuating coupling \cite{nguyen_high-coherence_2019}. Similarly, the main source of infidelity in semiconductor spin qubits is charge noise, which also causes fluctuations in coupling strength \cite{van_dijk_impact_2019}. However, the main cause of infidelity for fast gating of transmon qubits is leakage \cite{motzoi_simple_2009}. Therefore, in this paper, we show how to use this decomposition to create a universal set of gates that is either robust to coupling fluctuations or leakage for any commonly used qubit array structure.

This work is structured as follows. First in Sec.~\ref{sec:decomp}, we explain the method for decomposing the arrays into two-edge, three-edge, and four-edge vertices. In Sec.~\ref{sec:numericalmethod} we describe the numerical method for creating gates robust to coupling fluctuations or leakage. Then in Sec.~\ref{subsec:robustJ} we present a universal set of gates robust to fluctuations in coupling for each type of vertex, while in Sec.~\ref{subsec:leakage} we present a universal set of gates robust to leakage. We summarize and conclude in Sec.~\ref{sec:conclusions}.

\section{Hamiltonian}\label{sec:decomp}
To make use of the Hamiltonian decomposition in Ref.~\cite{kanaar_non-adiabatic_2022}, Ising coupling is required for each type of qubit array. Therefore, this section starts with deriving Ising coupling for each type of qubit array. Then the general qubit Hamiltonian used for the optimization is found. Finally, the choice of universal gate set for this Hamiltonian is discussed.

\subsection{Ising coupling for semiconductor spin qubits}
Ising coupling can be directly implemented in a semiconductor spin qubit device through the exchange interaction. The exchange interaction is the simplest two-qubit interaction in these devices and comes from the overlap of the electron wave functions of neighboring qubits \cite{burkard_semiconductor_2021}. Next-nearest neighbor interactions are negligible. The Hamiltonian terms that result from the exchange interaction have the form $\frac{J}{4} (XX+YY+ZZ)$, which can be derived using a Schrieffer–Wolff transformation on the Hubbard model \cite{loss_quantum_1998,kanaar_single-tone_2021}. After applying the rotating wave approximation (RWA) in the presence of a large Zeeman energy difference (i.e., a Hamiltonian term of the form $ZI-IZ$), only the $ZZ$ term is left. Single qubit driving is usually achieved through electron spin resonance (ESR) or electron dipole spin resonance (EDSR) but other novel methods exist as well \cite{benito_electric-field_2019,gilbert_-demand_2023}. Additionally, we assume the driving phase can be changed freely such that virtual-$z$ gates can be implemented \cite{mckay_efficient_2017}.

The largest source of infidelity in silicon semiconductor spin qubits is charge noise. Charge noise causes voltage fluctuations which results in $1/f$ frequency-dependent fluctuations in the exchange $J$\cite{connors_low-frequency_2019}. 
Gates robust to quasistatic $J$ fluctuations also performs well against $1/f$ frequency-dependent $J$ fluctuations. Therefore, a set of gates robust to quasistatic $J$ fluctuations will be optimized in this work.

\subsection{Ising coupling for transmon or flux qubits}\label{sec:couplingsuperconducting}
Although different types of couplings result in Ising coupling between superconducting qubits, this section derives $ZZ$ interaction for inductive coupling because of its simplicity. Inductive coupling can be modeled by Hamiltonian terms proportional to $\sin(\phi_1+\varphi)\sin(\phi_2+\varphi)$ where $\phi_i$ is the phase of the $i$-th qubit and $\varphi$ is the external flux \cite{rasmussen_superconducting_2021}. At $\varphi=\frac{\pi}{2}$ this coupling is $-(\phi_1^2+\phi_2^2)+\phi_1^2 \phi_2^2+1/144(\phi_1^4+\phi_2^4)$ if expanded to fourth order in $\phi_i$. The $\phi_i^2$ and $\phi_i^4$ terms are an effective adjustment of the resonant frequencies and anharmonicities respectively. The $\phi_1^2 \phi_2^2$ term results in Ising coupling in the logical subspace. Additionally, using the RWA the inductive coupling between leakage states is negligible in the frame rotating with the qubit frequencies. Ising coupling between superconducting qubits can also be created by using a transmon element as a coupler \cite{collodo_implementation_2020}. Alternatively, Ising coupling can be derived by applying a Schrieffer-Wolff frame transformation to direct capacitive or resonator-mediated coupled superconducting qubits \cite{long_universal_2021}. However, single qubit driving in the original frame of that method ends up driving both qubits in the transformed frame. Therefore, in that case, the driving method and device parameters would need to be carefully tuned such that the driving in the transformed frame is on one qubit and the decomposition from Sec.~\ref{sec:decomp2} still works.

The main source of infidelity for flux qubits is fluctuations in the magnetic flux called flux noise. Fluctuations in the magnetic field result in fluctuations of the coupling strength, $J$, when using inductive coupling. Additionally, these fluctuations are approximately $1/f$ frequency-dependent \cite{bylander_noise_2011}. This means that a solution that is robust against quasistatic $J$ fluctuations will perform well against both flux noise and charge noise. For this reason, we only need to theoretically find one set of gates robust to $J$ fluctuations, which can then be applied to either superconducting flux qubits or silicon semiconductor spin qubits.

The main source of infidelity for transmon qubits is leakage. Taking the first leakage level into account leads to modified decomposition in terms of a set of $\mathfrak{su}$(3) Hamiltonians. For the modified decomposition, we design a set of gates robust to leakage rather than $J$ fluctuations. 

\begin{figure}
    \centering
    \includegraphics[width=0.45\linewidth]{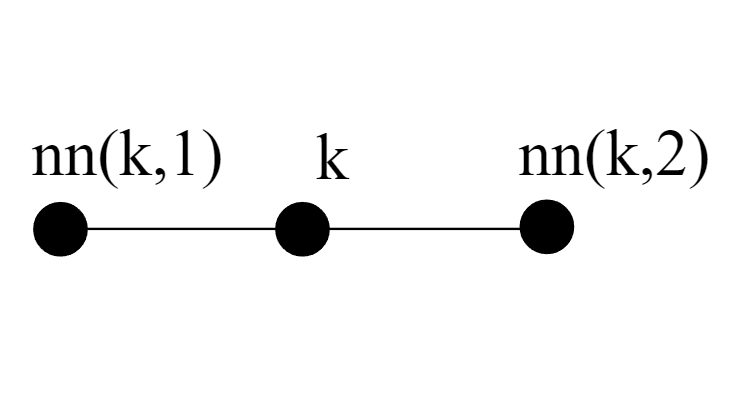}
    \includegraphics[width=0.4\linewidth]{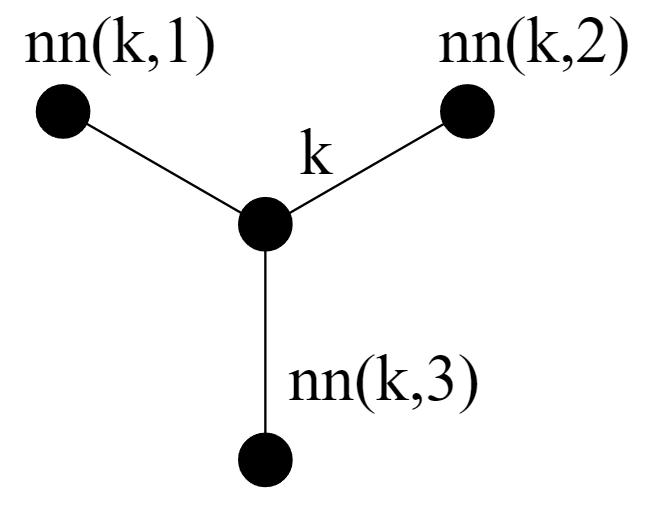}
    \includegraphics[width=0.5\linewidth]{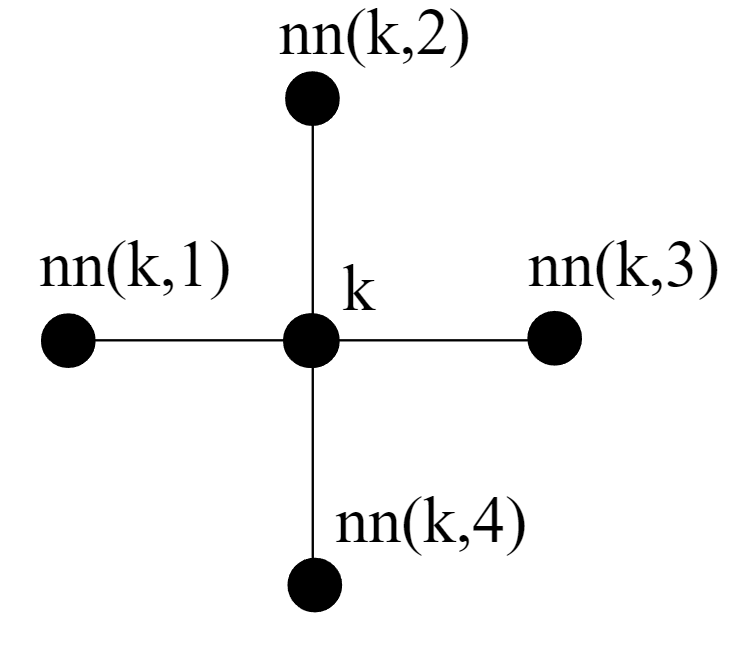}
    \caption{The two-edge (left), three-edge (right), and four-edge (bottom) vertices of which arrays are composed. $k$ indicates the driven center qubit while the lines indicate the fixed interaction. The center qubit's $i$-th nearest neighbor qubit is labeled with $nn(k,i)$.}
    \label{fig:intersections}
\end{figure}
\begin{figure}
    \centering
    \includegraphics[width=0.8\linewidth]{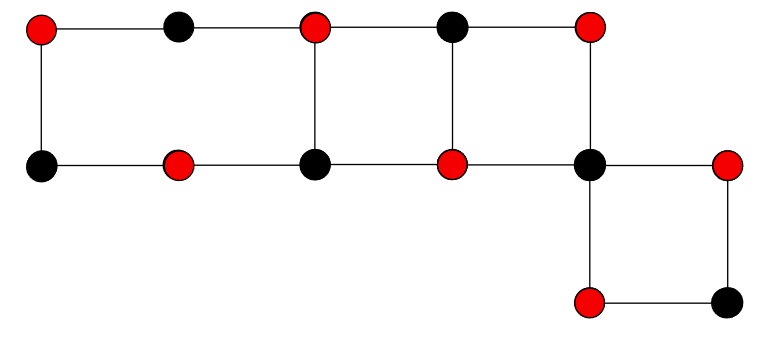}
    \caption{Generic qubit array represented by dots where the coupling between qubits is indicated by lines. Two sets of qubits needed for the decomposition are shown in black and red. The Hamiltonian decomposes into sets of commuting $\mathfrak{su}$(2) subalgebras when driving all red or black qubits.}
    \label{fig:qubitsets}
\end{figure}
\subsection{Types of vertices}
We represent a qubit array in terms of a graph of vertices (qubits) and undirected edges (coupling links). Any of the standard 2D arrays currently used for quantum computing can be represented using only two-edge, three-edge, and four-edge vertices, as depicted in Fig.~\ref{fig:intersections}. The quantum computing devices IBM has produced with transmon qubits \cite{kim_evidence_2023} can be constructed using the two-edge and three-edge vertices. The square qubit arrays composed of semiconductor spin qubits \cite{unseld_2d_2023,philips_universal_2022,lawrie_simultaneous_2023} or superconducting qubits \cite{acharya_suppressing_2023} can be constructed from two-edge vertices at the corners, three-edge vertices on the edges and four-edge vertices for all other qubits. Therefore, it is sufficient to show how to create a universal gate set within these three of types of vertices for universal control of qubit arrays. 

\subsection{Hamiltonian decomposition}\label{sec:decomp2}
The decomposition described in this section applies to any qubit array with $ZZ$ interactions between nearest neighbors with the extra condition that the set of nearest neighbors is disjoint from the set of next nearest neighbors for every qubit in the array. Additionally, we assume single qubit control is possible for every qubit as it is necessary for universal quantum computation. The Hamiltonian, $H$, for this system is
\begin{equation}
    H=\sum_{i=1}^{N}\frac{\Omega_i}{2}(\cos(\phi_i)X_i+ \sin(\phi_i)Y_i)+\frac{1}{2}\sum_{j=1}^{n_i} \frac{J_{nn(i,j)}}{4} Z_i Z_{nn(i,j)} 
\end{equation}
where $N$ is the number of qubits, $\Omega_i$ and $\phi_i$ are the driving strength and phase on the $i$-th qubit, $X_i$ is the $x$ Pauli operator on the $i$-th qubit, $n_i$ is the number of nearest neighbors of the $i$-th qubit, $nn(i,j)$ is the $j$-th nearest neighbor of the $i$-th qubit and $J_{nn(i,j)}$ is the coupling strength between the $i$-th qubit and its $j$-th nearest neighbor. Since the coupling terms are counted twice, there is a factor of half in front of the coupling sum. Additionally, the coupling coefficient $J/4$ is conventional for semiconductor spin exchange interaction whereas in the superconducting context, this would typically be denoted as simply $J$. As stated at the beginning of this section, we are considering qubit arrays that can be decomposed into two sets of qubits where no two qubits in a given set are directly coupled. Two such sets are shown in Fig.~\ref{fig:qubitsets}. By locally driving only within one of these sets at a time one ensures that the part of the Hamiltonian involving operators on a given qubit commute with the rest of the Hamiltonian. The whole Hamiltonian is thus a sum of mutually commuting terms, 
\begin{equation}
    H=\sum_k H_k,
\end{equation}
with
\begin{equation}
    H_k = \frac{\Omega_k}{2} (\cos (\phi_k)X_k+\sin (\phi_k)Y_k)+ \sum_{j=1}^{n} \frac{J_{nn(k,j)}}{4}Z_k Z_{nn(k,j)}, \label{Eq:HamDecomp1}
\end{equation}
where $k$ runs only over the vertices in the driven set. The $k$th term further decomposes into a set of $2^{n}$ $\mathfrak{su}$(2) Hamiltonians where $n$ is the number of neighbors the $k$th vertex has. The resulting decomposition into $\mathfrak{su}$(2) Hamiltonians was shown in Ref.~\cite{kanaar_non-adiabatic_2022} for two and three neighbors and that result generalizes to
\begin{equation}\label{eq:Hk}
    H_k = \sum_{\vec{s}} H_{k, \vec{s}}, 
\end{equation}
where $\vec{s}=\left(s_1,s_2,...,s_n\right)$, $s_i \in \{+,-\}$,
\begin{multline}\label{eq:Hks}
    H_{k, \vec{s}} =  \frac{1}{4} \sum_{i=1}^{n} s_i J_{nn(k,i)} Z_{k,\vec{s}}
    \\
    +\frac{\Omega_k}{2} \left(\prod_{i=1}^{n} s_i\right)\left(\cos (\phi_k)X_{k,\vec{s}}+\sin (\phi_k)Y_{k,\vec{s}}\right),
\end{multline}
and we have defined 
\begin{equation}\label{eq:Xks}
X_{k,\vec{s}}=\frac{1}{2^n}X_k\prod^n_{i=1}(Z_{nn(k,i)}+s_i I_{nn(k,i)}),   
\end{equation}
and likewise for $Y$ and $Z$.

For transmon qubits, including the first excited leakage state, $\ket{2}$, of each qubit slightly modifies Eq.~\eqref{eq:Hks} into the form
\begin{multline}\label{eq:HksTransmon}
        H_{k,\Vec{s}} = \frac{1}{4}\sum_{i=1}^n s_i J_{nn(k,i)} Z_{k,\Vec{s}} + \frac{\Delta_k}{2} (I^L_{k,\vec{s}}-Z^L_{k,\vec{s}})
        \\
        +\frac{\Omega_k}{2}\left(\prod_{i=1}^n s_i\right) \left[ \cos (\phi_k)\left(X_{k,\Vec{s}}+\lambda_2 X^L_{k,\Vec{s}}\right) \right.
        \\
        \left. + \sin (\phi_k) \left(Y_{k,\Vec{s}} + \lambda_2 Y^L_{k,\Vec{s}} \right) \right] ,
\end{multline}
where $\lambda_2$ is the ratio of the driving strength between the $\ket{1}\leftrightarrow\ket{2}$ transition and the $\ket{0}\leftrightarrow\ket{1}$ transition (which later will be set to $\lambda_2=1$ for the optimization), $X^L_{k,\Vec{s}}$ is defined as in Eq.~\eqref{eq:Xks} except that it acts on the subspace formed by $\{\ket{1},\ket{2}\}$ instead of $\{\ket{0},\ket{1}\}$, and $\Delta$ is the anharmonicity. 

Despite the always-on $ZZ$ couplings, we can use this decomposition to design a pulse that individually controls any qubit.
This decomposition also works for any coupling that can be written as a direct product of local operators, excluding identities.

\subsection{Universal gate set}\label{sec:universalgates}
A universal set of gates consisting of $\frac{\pi}{2}$ $ZZ$ rotations, $X$-gates, and arbitrary angle $Z$ rotations are presented in this work. Arbitrary angle $Z$ rotations can be implemented virtually \cite{mckay_efficient_2017}. Non-robust $X$-gates can be implemented analytically by using square pulses sequences for two-edge and three-edge vertices as shown in Ref.~\cite{kanaar_non-adiabatic_2022}. However, designing $X$-gates robust to $J$ fluctuations or leakage analytically would be very challenging, so in this work we find them numerically with smooth control pulses as described in Sec.~\ref{sec:numericalmethod}. $ZZ$ rotations between two specific qubits cannot be directly implemented in the decomposed Hamiltonian since all couplings are on at the same time. Nonetheless, it is possible to isolate a particular $ZZ$ coupling by echoing out any unwanted $ZZ$ rotations with $X$-gates on the untargeted nearest neighbor qubits.

In addition to performing the desired gate on a target qubit, it is also necessary for the other qubits to remain in the same state. In the absence of driving, idle qubits entangle through $\sum_{j=1}^n Z_k Z_{nn{(k,i)}}$ rotations. In the absence of $J$ fluctuations, it is possible in principle to time the gates such that all the $\sum_{j=1}^n Z_k Z_{nn{(k,i)}}$ evolution is a multiple of a local $\pi$ $ZZ$ rotation which can be reversed using virtual-$z$ gates. This method is even possible in the presence of leakage because inductive coupling does not cause leakage. However, this is only practical if the couplings in the array take only a small set of discrete values. Furthermore, in the presence of $J$ fluctuations free evolution of the coupling would lead to decoherence.

A more general approach is to create an identity gate by performing an $X$-gate in the middle and at the end of the time for which the qubit is desired to remain idle. This is because any $ZZ$ rotation in the first half would be canceled by the echoed opposite sign rotation in the second half. Further, if a qubit needs to be idle for a long time it might be beneficial for many echos to be performed in a manner similar to  Carr-Purcell-Meiboom-Gill (CPMG) dynamical decoupling pulses \cite{carr_effects_1954,meiboom_modified_1958}. Additionally, the idle time needs to be at least as long as the time to perform the two $X$-gates that implement the identity gate. Alternatively, it is possible to find a robust identity gate that takes the same time as the $\frac{\pi}{2}$ $\sum_{j=1}^n Z_k Z_{nn{(k,i)}}$ rotation and $X$-gates. This is the approach taken in Sec.~\ref{subsec:robustJ} for pulses robust to $J$ fluctuations.\par
For pulses robust to leakage only $X$-gates were optimized because the $Z$ gates can be done virtually and the $\frac{\pi}{2}$ $ZZ$ rotations and identity gates can be implemented using $X$-gate echos. Therefore, for transmons, only $X$-gates robust to leakage were optimized for each vertex. However, for flux and semiconductor spin qubits, $X$-gates, identity gates, and  $\frac{\pi}{2}$ $\sum_{j=1}^n Z_c Z_{c,nn{(c,i)}}$ rotations robust to coupling fluctuations were optimized for each type of vertex.

\section{Numerical optimization method} \label{sec:numericalmethod}
We find gates robust against $J$ fluctuation or leakage by numerically minimizing a cost function consisting of three terms,
\begin{equation}\label{eq:C}
    \mathcal{C} = \mathcal{C}_{\text{gate}} + \mathcal{C}_{\text{robust}} +\mathcal{C}_{\text{constraint}}.
\end{equation} 
The first term guides the gate produced to minimize its trace infidelity with the desired target gate. Because of the Hamiltonian decomposition, this term can be broken down into the projections of the targeted gate into the distinct $SU(2)$s generated by the $\mathfrak{su}(2)$ subalgebras of Eq.~\eqref{eq:Hks}. Although there are a daunting $2^n$ such terms for a vertex with $n$ neighbors, most of them are identical in the symmetric case $J_{nn(k,i)}=J$, leaving only $N_p$ distinct $SU(2)$ evolutions, where $N_p = 3$, 4, or 5 for a two-edge, three-edge, or four-edge vertex respectively. Thus, we choose the cost function to be the weighted average of the individual $SU(2)$ trace infidelities, 
\begin{equation}\label{eq:Ctrace}
    \mathcal{C}_{\text{gate}} =  \sum_{i=1}^{N_p} w_i \left[1-\left|\frac{1}{2}\text{Tr} (U_{c,i} U_{t,i}^{\dagger})\right|^2\right],
\end{equation}
where $U_{c,i}$ is the gate resulting from the control pulse in question in the absence of error, projected onto the $i$th $SU(2)$, and $U_{t,i}$ is the target gate, also projected onto the same $SU(2)$. $w_i$ is the number of times the $i$-th distinct $SU(2)$ evolutions occurs in the full evolution.

To make a gate robust against quasistatic $J$ fluctuations we use the first-order Magnus expansion. 
An error in the physical coupling, $\delta J_{nn(k,i)}$, results in errors on all the decomposed Hamiltonians of Eq.~\eqref{eq:Hks}. For a small $\delta J_{nn(k,i)}$ the gate, $U_i$, can be approximated as 
\begin{equation}
    U_i \approx U_{c,i} e^{ i \mathcal{E}_i}.
\end{equation}
$\mathcal{E}_i$ is the integral of the error Hamiltonian $H_{\epsilon,i}=\delta J_{\vec{s}} Z_{\vec{s}}$ in the toggling frame, i.e. $\mathcal{E}_i =\int_0^T U_{c,i}^{\dagger} H_{\epsilon,i} U_{c,i} dt$. If $\mathcal{E}_i$ is minimized, the gate is robust to errors $\delta J_{\vec{s}}$. In our case, the error Hamiltonians, $J_{\Vec{s_i}} Z_{\Vec{s_i}}$, appear within each $\mathfrak{su}$(2). Therefore, to create a gate robust against $J$ fluctuations, the second term in the cost function is the weighted sum of the square of the Frobenius norms of $\mathcal{E}_i$,
\begin{equation}\label{eq:Crobust}
    \mathcal{C}_{\text{robust}} =0.3 \sum_{i=1}^{N_p} w_i \left| \int_0^{JT} U_{c,i}^{\dagger}(\frac{\tau}{J})  Z_{\Vec{s_i}} U_{c,i}(\frac{\tau}{J}) d\tau \right|,
\end{equation} 
where we empirically found the 0.3 weighting to give a good compromise between noiseless gate fidelity \eqref{eq:Ctrace} and robustness \eqref{eq:Crobust} in Eq.~\eqref{eq:C}.
To account for leakage in transmon qubits, we use the expanded $\mathfrak{su}(3)$ Hamiltonians from Eq.~\eqref{eq:HksTransmon} which include the first leakage level. As stated in Sec.~\ref{sec:couplingsuperconducting}, inductive coupling does not couple different subspaces to each other. This means that the gate is robust to leakage as long as the gate in the logical subspace, $\{\ket{0},\ket{1}\}$, is the desired gate. Since the infidelity in Eq.~\eqref{eq:Ctrace} is defined for the logical space, optimizing it leads to a leakage robust gate.

We optimize the driving strength, $\Omega(t)$, and phase, $\phi(t)$, via the parameterization
\begin{equation} \label{Eq:sinseriesansatz}
\begin{split}
\Omega(t) \cos(\phi(t)) & = \sum_{n=1}^{m/2} a_n \sin \left(\frac{n \pi t }{T} \right) \\
\Omega(t) \sin(\phi(t)) & =\sum_{n=m/2+1}^{m} a_n \sin \left(\frac{n \pi t }{T} \right) 
\end{split}
\end{equation}
In Eq.~\eqref{Eq:sinseriesansatz} there are $m$ free parameters, $a_n$, split over the in-phase and quadrature parts of the driving field. This parameterization intrinsically respects the physical considerations of a limited bandwidth and a driving strength that starts and ends at zero.
However, the physical constraint on available driving strength is not built in, so have included it via the cost function,
\begin{equation}
    \mathcal{C}_{\text{constraint}} =0.003\frac{1}{T}\int_0^T \text{relu}\left(1-\left(\frac{\Omega(t)}{\Omega_{\text{max}}}\right)^2\right)dt,
\end{equation}
where relu is the rectified linear unit function and we empirically found the 0.003 weighting to achieve a good fidelity and robustness while staying below the maximum driving strength.
The evolution operator for each Hamiltonian was numerically solved in Julia with the BS5 solver from the DifferentialEquations.jl package. The cost function was minimized using the BFGS algorithm from the Optim.jl package.

\section{Results}\label{sec:results}
The pulses presented in this section were optimized to be either robust to leakage or $J$ fluctuations for each type of vertex. We report the fastest optimal pulses we found for the two-, three- and four-edge vertices. Similar performance can also be obtained for longer pulse times, and when dealing with a qubit array consisting of multiple types of vertices all gates could be performed in parallel by using the same gate time for all the vertices, in which case that time would be determined by whichever vertex has the most neighbors.

\subsection{Gates robust to $J$ fluctuations}\label{subsec:robustJ}
Here we consider systems where the main cause of infidelity is fluctuations of the interaction strength $J$. This includes electron spin qubits in silicon quantum dots (whether Si/SiO$_2$ or Si/SiGe heterostructures), where the main source of infidelity is charge noise, as well as fluxonium qubits, where the main source of infidelity is flux noise. The pulse times for the two-edge, three-edge, and four-edge vertices are $T=4.5\pi/J$, $T=6\pi/J$, and $T=10\pi/J$ respectively. The maximum driving strength used in this section, $\Omega_{max}=0.4 J$, is consistent with semiconductor spin qubit experimental values \cite{philips_universal_2022,mills_two-qubit_2022,huang_high-fidelity_2023} while Ref.~\cite{le_scalable_2023} used a much larger maximum drivings strength relevant to superconducting qubits. The results in this section can also be used for flux qubits by choosing $J$ appropriately.  

The optimized weights of the sine series parameterizing $\Omega \cos (\phi)$ and $\Omega \sin (\phi)$ are reported in the supplementary material. The optimization for the identity gate and $\frac{\pi}{2}$ $ZZ$ rotations for every type of vertex resulted in nearly identical in-phase and quadrature components of the driving field, effectively driving the $\frac{X+Y}{\sqrt{2}}$ axis. The orientation of the $X$ and $Y$ axes does not matter for these gates therefore by changing the driving phase by $\frac{\pi}{4}$ these gates can be performed with single-axis driving. The highest frequencies used in the optimized gates for the two-edge, three-edge, and four-edge vertices are $40/(9 \pi)J \approx 1.5 J$, $30/(12 \pi)\approx 0.8 J$ and $50/(20 \pi)J\approx 0.8 J$. For $J\sim 10$MHz these frequencies are significantly lower than the $\sim 100$MHz low-pass filters used in experiments \cite{xue_quantum_2022,xue_cmos-based_2021}.

\begin{figure}
    \centering
    \includegraphics[width=0.85\linewidth]{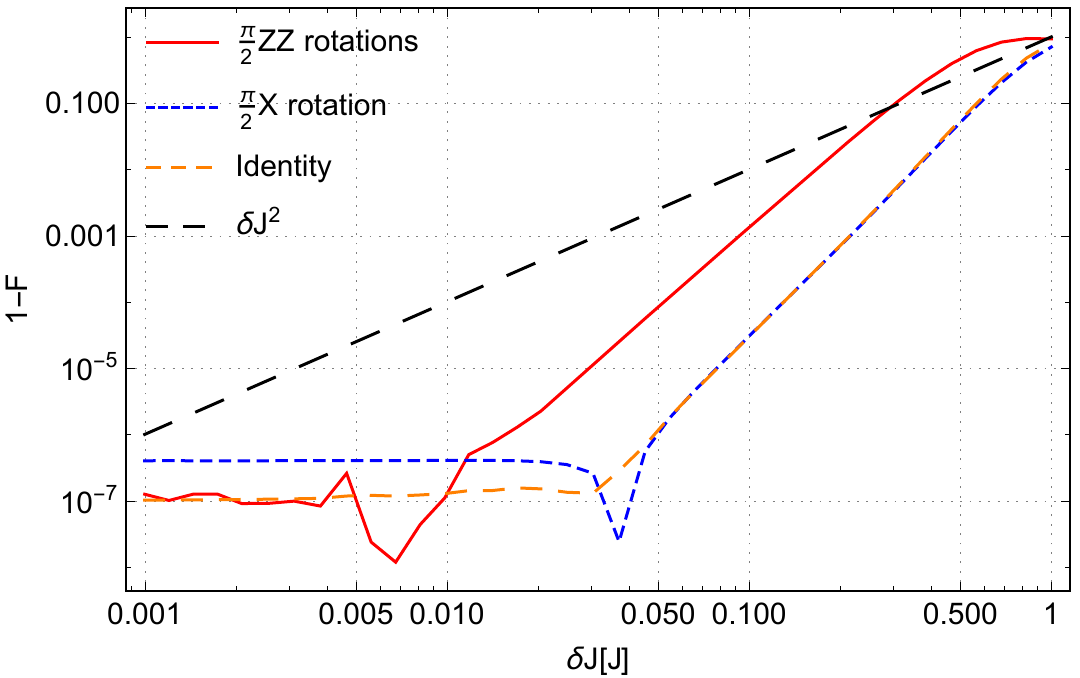}
    \includegraphics[width=0.85\linewidth]{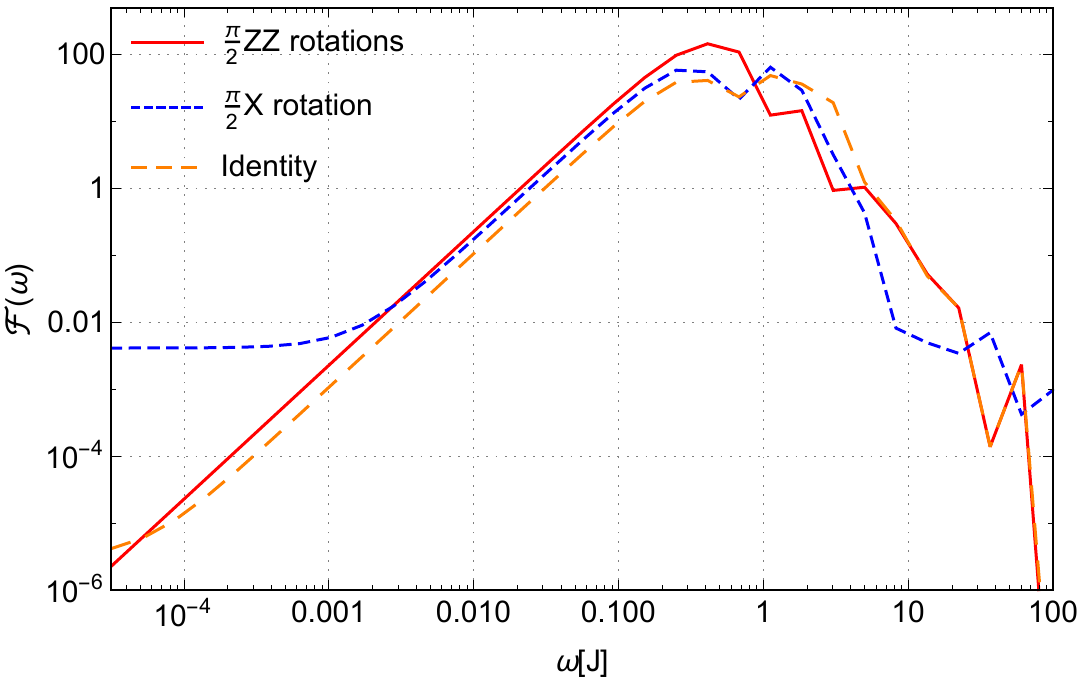}
    \caption{Top: Infidelity of the $\frac{\pi}{2}$ $ZZ$ rotation, $X$-gate, and identity gate for a chain vertex as a function of quasistatic fluctuations in $J$. Bottom: Filter function of the same gates as a function of frequency $\omega$.}
    \label{fig:ChainJFid}
\end{figure}

\begin{figure}
    \centering
    \includegraphics[width=0.85\linewidth]{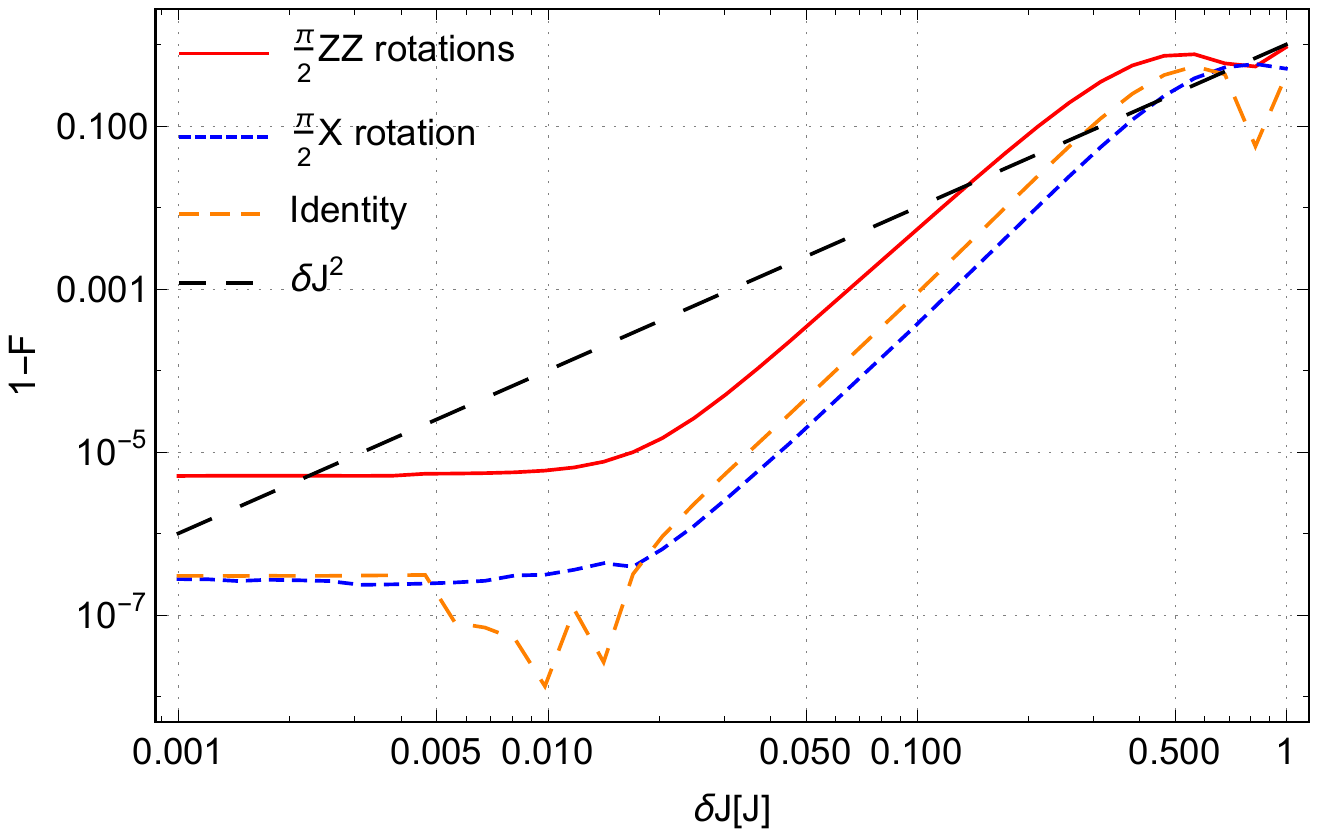}
    \includegraphics[width=0.85\linewidth]{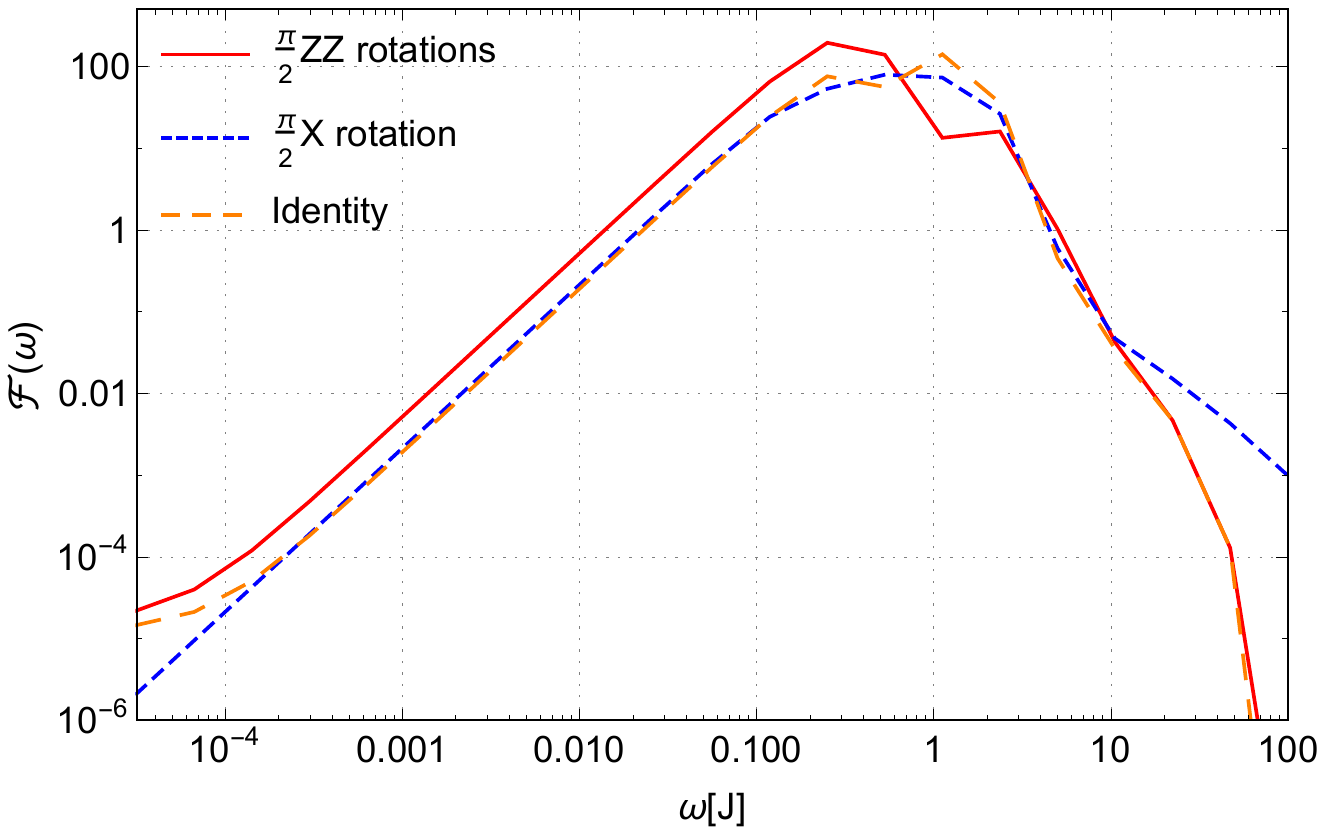}
    \caption{Top: Infidelity of the $\frac{\pi}{2}$ $ZZ$ rotation, $X$-gate, and identity gate for a honeycomb vertex as a function of quasistatic fluctuations in $J$. Bottom: Filter function of the same gates as a function of frequency $\omega$.}
    \label{fig:HonJFid}
\end{figure}

\begin{figure}
    \centering
    \includegraphics[width=0.85\linewidth]{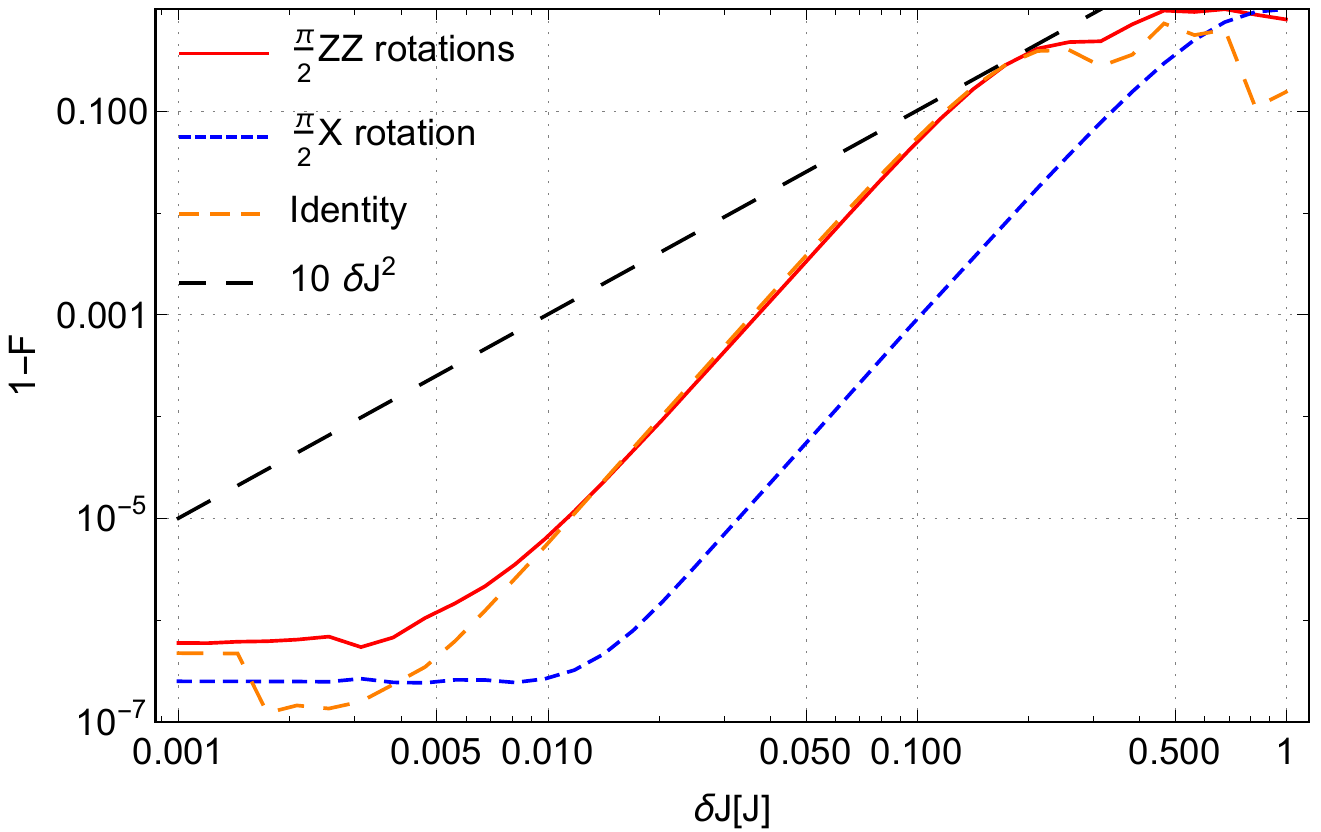}
    \includegraphics[width=0.85\linewidth]{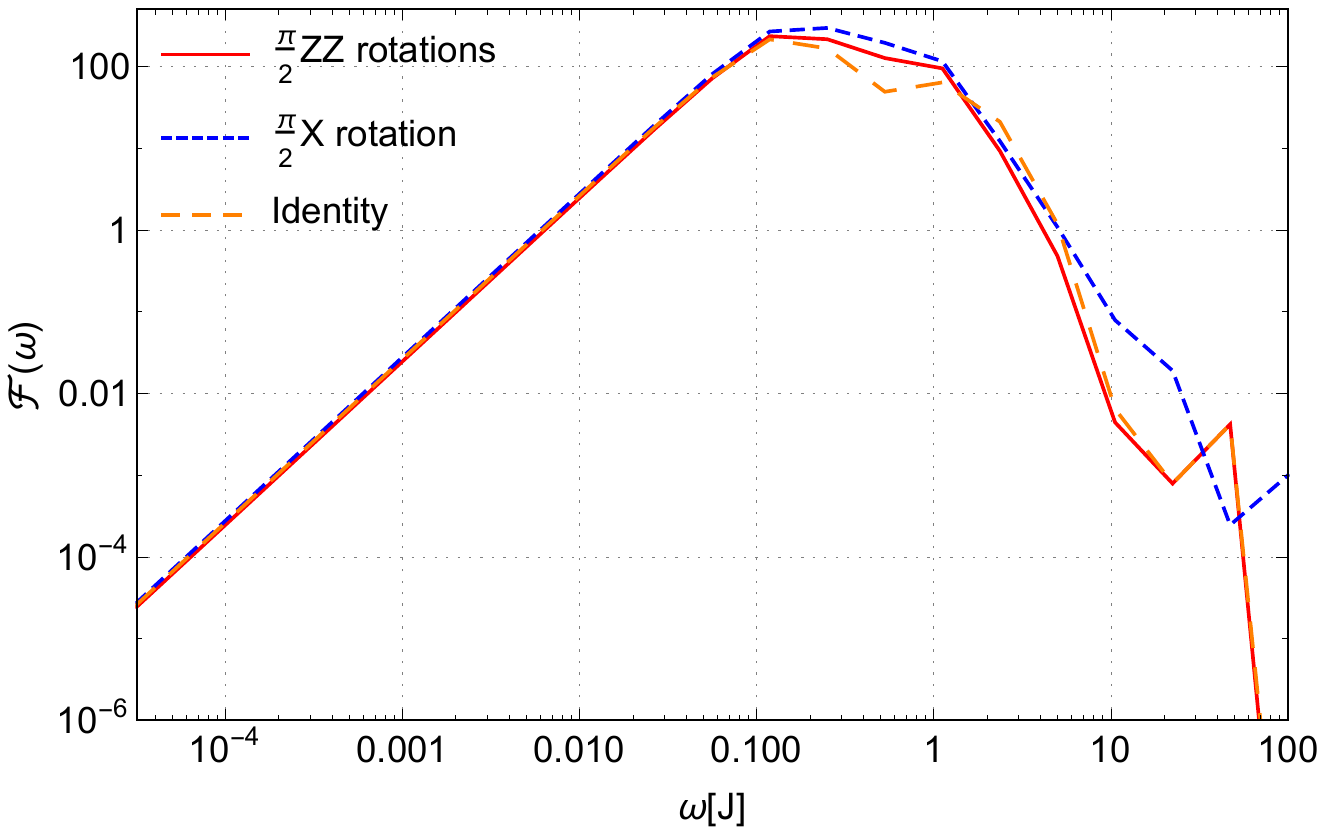}
    \caption{Top: Infidelity of the $\frac{\pi}{2}$ $ZZ$ rotation, $X$-gate, and identity gate for a square vertex as a function of quasistatic fluctuations in $J$. Bottom: Filter function of the same gates as a function of frequency $\omega$.}
    \label{fig:SquareJFid}
\end{figure}

The trace infidelity of the $n+1$ qubit gates, $1-F$, versus quasistatic fluctuations of $J$ for the different gates are shown in Figs.~\ref{fig:ChainJFid}-\ref{fig:SquareJFid} for two-edge, three-edge, and four-edge vertices respectively. All these gates reach an infidelity below $10^{-5}$ in the absence of any fluctuations and scale better than $\sum_i^n \delta J_{nn(k,i)}^2$ versus quasistatic fluctuations.

However, in reality, charge and flux noise are not merely quasistatic but have a roughly $1/f$ frequency dependence. To analyze the effect of frequency-dependent noise on the gate infidelities, the filter function formalism is used \cite{green_arbitrary_2013}. The filter function, $\mathcal{F}(\omega)$, is the Fourier transform of the physical error Hamiltonians, $\frac{J_{nn(k,i)}}{4} Z_kZ_{nn(k,i)}$, in the interaction frame
\begin{equation}
    \mathcal{F}(\omega)=\sum_{i=0}^n  \sum_{j=0}^{4^{n+1}} \frac{J_{nn(k,i)}}{4 \times 2^{n+1}}\text{Tr}\left(\sigma_j \int_{0}^{T} e^{i \omega t} U_c^{\dagger}   Z_kZ_{nn(k,i)} U_c dt\right)
\end{equation}
where $\sigma_i$ is the $i$-th basis matrix of SU($2^{n+1}$) formed from Kronecker products of Pauli matrices. The filter functions as a function of frequency for the different vertices are shown in Figs.~\ref{fig:ChainJFid}-\ref{fig:SquareJFid}. The infidelity as a result of a noise spectral density, $S(\omega)$, is calculated with  $1-F=\frac{1}{2\pi}\int^{\infty}_{\omega_{\text{ir}}} S(\omega)\mathcal{F}(\omega)\text{d}\omega$. We take the power spectral density of charge or flux noise to have the form
\begin{equation}
    S(\omega) = 
  \begin{cases}
    A_0^2/\omega  & \text{for } \omega_{\text{ir}} \leq \omega \leq \omega_{\text{cutoff}} \\
    A_0^2\omega_{\text{cutoff}}/\omega^2 & \text{for } \omega_{\text{cutoff}} \leq \omega \leq \infty
  \end{cases}.
\end{equation} 
Using a calibration cutoff $\omega_{\text{ir}}=10^{-3}$Hz, a cutoff frequency $\omega_{\text{cutoff}}=100$MHz and a $1$Hz variance $A_0=3\mu \text{V}/\sqrt{\text{Hz}}$ for charge noise results in an infidelity below $10^{-3}$ for all gates for each vertex \cite{connors_low-frequency_2019}.

\subsection{Gates robust to leakage} \label{subsec:leakage}
\begin{figure}
    \centering
    \includegraphics[width=0.85\linewidth]{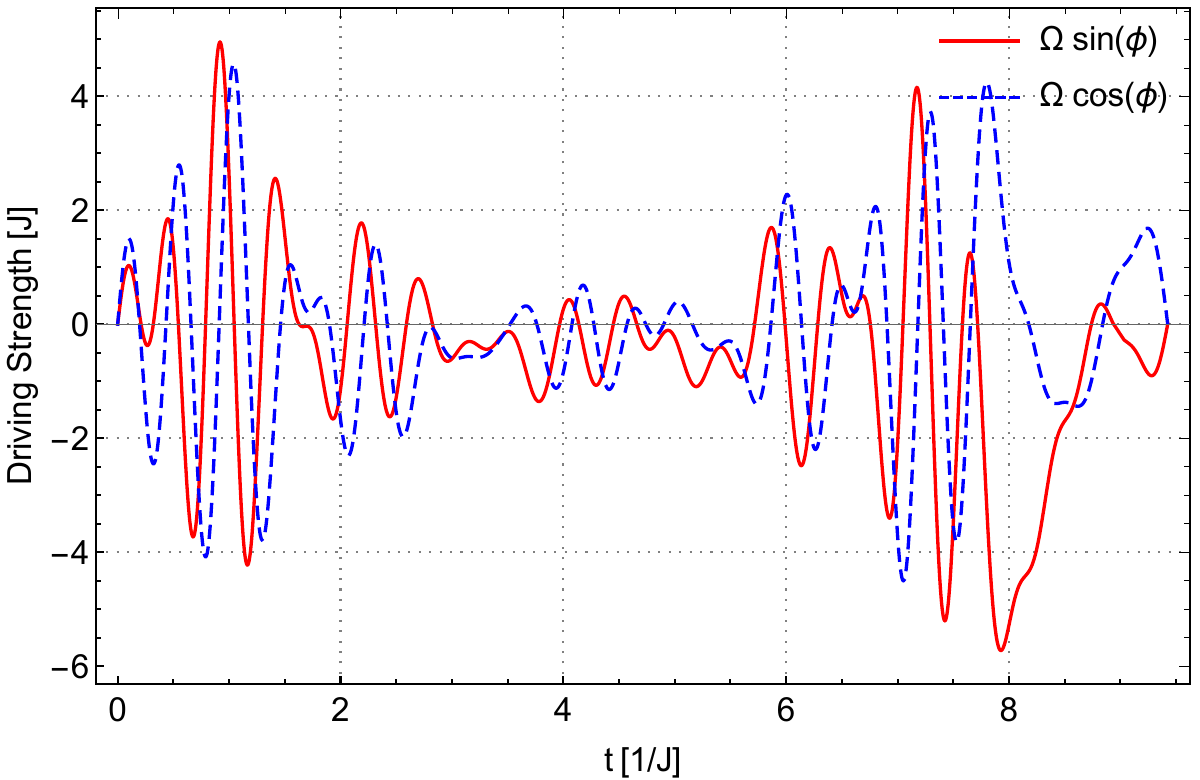}
    \includegraphics[width=0.85\linewidth]{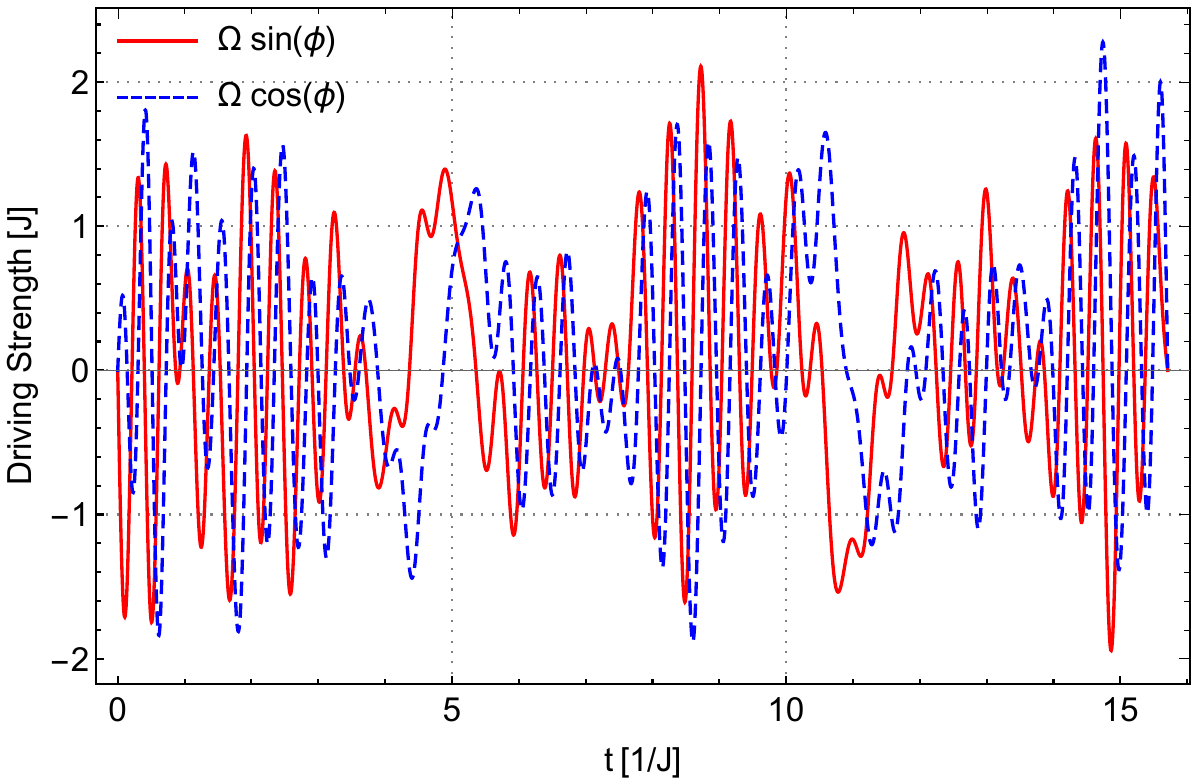}
    \includegraphics[width=0.85\linewidth]{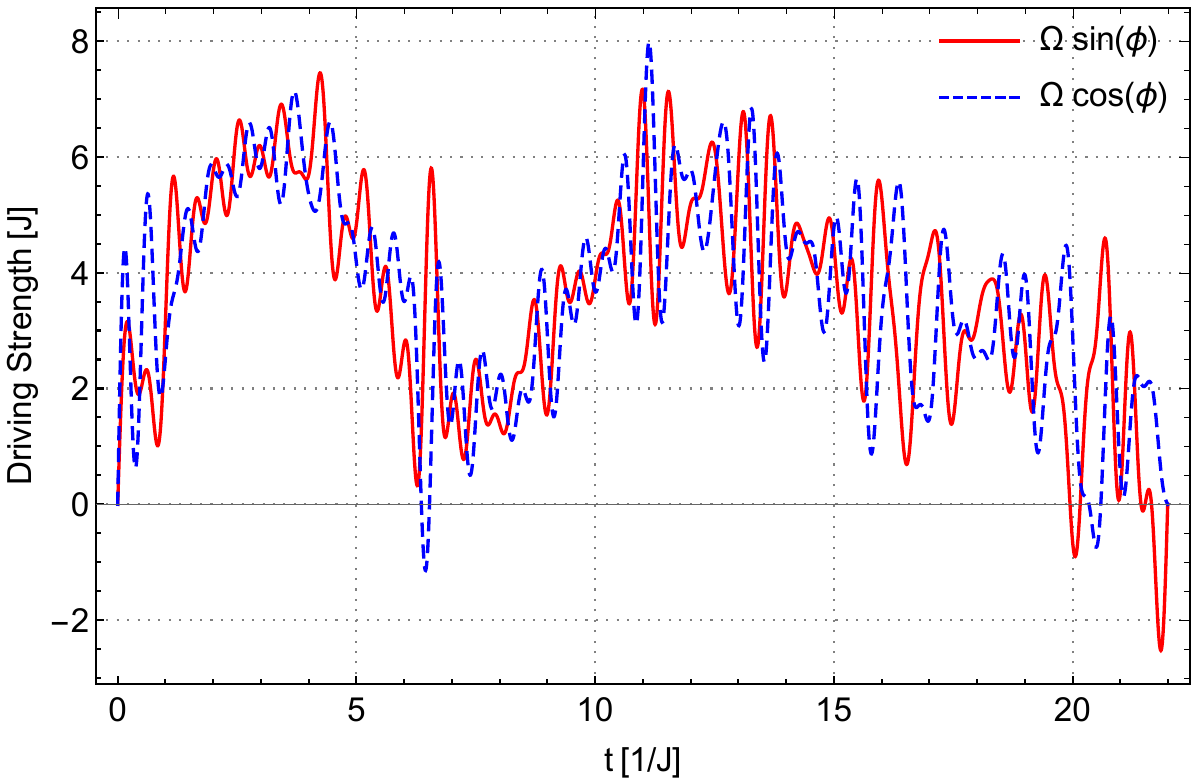}
    \caption{Driving fields on the $X$, $\Omega \cos(\phi)$ and $Y$ operators for $X$-gates that compensate for leakage in a two-edge vertex(top), three-edge vertex(middle) and four-edge vertex(bottom).}
    \label{fig:Leakagepulseshapes}
\end{figure}
As explained in Sec.~\ref{sec:universalgates}, when leakage is the dominant source of error, we only need to find a robust $X$-gate for all three vertex types.

The device parameters used in this section are an anharmonicity of $\Delta =300$MHz, an inductive coupling strength of $J=30$MHz, and a maximum driving strength of $\Omega_{\text{max}}=300$MHz. These values have been reached in experiment \cite{ciani_microwave-activated_2022}. The optimized pulse shapes resulting in $X$-gates are shown in Fig.~\ref{fig:Leakagepulseshapes}. The $X$-gates reach an infidelity below $10^{-7}$ for each type of vertex. However, the infidelities are very dependent on the value of the anharmonicity. For a miscalibration in the anharmonicity of $5$kHz an infidelity of around $10^{-3}$ is reached for each of the optimized pulses. This would mean that the anharmonicity would need to be measured with an accuracy of $5$ kHz to maintain a good infidelity. If the anharmonicity is found by measuring the difference between the $\ket{0}\leftrightarrow \ket{1}$, $\ket{0}\leftrightarrow \ket{2}$, and $\ket{1}\leftrightarrow \ket{2}$ transition frequencies, which are in the GHz range \cite{schreier_suppressing_2008}, a resonator with a quality factor $Q=5 \times 10^5$ \cite{muller_magnetic_2022,samkharadze_high-kinetic-inductance_2016} would be needed to achieve a $5$ kHz accuracy in measuring the anharmonicity. The highest frequency used in the two-edge, three-edge, and four-edge vertex pulses is $50/(6 \pi)\approx 2.7 J$, $100/(10 \pi)\approx 3.2 J$, and $100/(14 \pi)\approx 2.3 J$ respectively. At $J=30$MHz these frequencies are below the Gaussian filter width of $225$MHz used on the controls in superconducting experiments \cite{dumur_v-shaped_2015,barends_superconducting_2014}. 

\section{Summary \& Conclusion} \label{sec:conclusions}
To help scale qubit devices, we investigated semiconductor spin and superconducting qubit arrays with fixed Ising coupling since they require less control wiring and calibration. Creating a universal set of gates for quantum computing within such an array is not trivial because of the fixed coupling. Therefore, we simplify the Hamiltonian by considering driving on a specific subset of qubits of such an array that results in a decomposition of the Hamiltonian into a set of commuting $\mathfrak{su}$(2) subalgebras for each driven qubit. We previously showed how to analytically create square pulse sequences that result in a universal set of gates for a chain or honeycomb array in Ref.~\cite{kanaar_non-adiabatic_2022}. Now, in this work, we numerically found a universal set of gates robust to either the main source of error in semiconductor spin and flux qubits, fluctuating coupling, or the main source of error in transmon qubits, leakage. In the latter case, we only needed to optimize the robustness of $X$-gates and the rest of the universal set of gates could be constructed by using these $X$-gates to form echo sequences for isolating specific interqubit coupling terms and by combining them with virtual-$z$ gates to perform local rotations. In contrast, in the presence of fluctuating coupling, we had to optimize robust $X$-gates, identity gates, and $\frac{\pi}{2}$ $\sum_{j=1}^n Z_c Z_{c,nn{(c,i)}}$ rotations. In all cases, we optimized pulses for two-edge, three-edge, and four-edge vertices, as shown in Fig.~\ref{fig:intersections}, so that our pulses can be used in any currently common qubit array structure. 

We have explicitly shown that even in a qubit array with fixed Ising coupling and only local control it is possible to create a universal set of robust, high-fidelity gates. This result shows one particular path towards fault-tolerant control of a scalable quantum computer.
 \section*{Acknowledgements} \label{sec:acknowledgements}
The authors acknowledge support from the National Science Foundation under Grant No. 1915064.

\bibliographystyle{apsrev4-1} 
\bibliography{refs} 

\end{document}